\documentclass[runningheads]{llncs}

\usepackage{graphicx}
\usepackage{amsmath, amssymb}
\usepackage{esvect}
\usepackage{hyperref}
\usepackage{multirow}
\usepackage{soul}
\usepackage{color}

\begin{document}

\title{Stress-testing cross-cancer generalizability of 3D nnU-Net for PET-CT tumor segmentation: multi-cohort evaluation with novel oesophageal and lung cancer datasets}



\author{
Soumen Ghosh$^{1,6\#}$, 
Christine Jestin Hannan$^{1,7\#}$, 
Rajat Vashistha$^{2,3*}$, 
Parveen Kundu$^{8}$, 
Sandra Brosda$^{3}$, 
Lauren G. Aoude$^{3}$,
James Lonie$^{4}$,
Andrew Nathanson$^{4}$,
Jessica Ng$^{4}$,  
Andrew P. Barbour$^{3,4,5}$, 
Viktor Vegh$^{2*}$ }
\institute{
$^{1}$ QIMR Berghofer Medical Research Institute, Brisbane, Australia \\
$^{2}$ Australian Institute for Bioengineering and Nanotechnology, The University of Queensland, Brisbane, Australia \\
$^{3}$ Frazer Institute, The University of Queensland, Brisbane, Australia \\
$^{4}$ Princess Alexandra Hospital, Brisbane, Australia \\
$^{5}$ PA Southside Clinical Unit, Faculty of Health, Medicine and Behavioural Sciences \\
$^{6}$ School of Biomedical Sciences, The University of Queensland, Brisbane, Australia\\
$^{7}$Department of Surgical Sciences, Uppsala University, SE-75185 Uppsala, Sweden\\
$^{8}$ Rohtak Nuclear Medcare Imaging Therapy and Research Center, India \\
*r.vashistha@uq.edu.au; v.vegh@uq.edu.au \quad \#Equal contribution
}

\maketitle

\begin{abstract}
Robust generalization is essential for deploying deep learning–based tumor segmentation in clinical PET-CT workflows, where anatomical sites, scanners, and patient populations vary widely. This study presents the first cross-cancer evaluation of nnU-Net on PET-CT, introducing two novel, expert-annotated whole-body datasets: 279 patients with oesophageal cancer (Australian cohort) and 54 with lung cancer (Indian cohort). These cohorts complement the public AutoPET dataset and enable systematic stress-testing of cross-domain performance. We trained and tested 3D nnU-Net models under three paradigms: target-only (oesophageal), public-only (AutoPET), and combined training. For the tested sets, the oesophageal only model achieved the best in-domain accuracy (mean DSC = 57.8) but failed on external Indian lung cohort (mean DSC $\leq$ 3.4), indicating severe overfitting. The public-only model generalized more broadly (mean DSC = 63.5 on AutoPET, 51.6 on Indian lung cohort) but underperformed in oesophageal Australian cohort (mean DSC =26.7). The combined approach provided the most balanced results (mean DSC = lung (52.9); oesophageal (40.7); AutoPET (60.9)), reducing boundary errors and improving robustness across all cohorts. These findings demonstrate that dataset diversity, particularly multi demographic, multi-center and multi-cancer integration, outweighs architectural novelty as the key driver of robust generalization. This work presents the demography based cross-cancer deep learning segmentation evaluation and highlights dataset diversity, rather than model complexity, as the foundation for clinically robust segmentation.

\keywords{Lesion Segmentation \and PET/CT \and nnU-Net \and Multi-Center Data}
\end{abstract}

\section{Introduction}
Positron Emission Tomography-Computed Tomography (PET-CT) provides complementary metabolic and anatomical information for the detection and delineation of tumors \cite{ahmad2025anatomy}. PET captures biological signals such as tissue metabolism, while CT provides detailed anatomical structure \cite{gatidis2022whole}\cite{vashistha2024modular}. Among various radiotracers, $^{18}$F-fluoro-deoxy-glucose (FDG) is the most widely used in PET imaging \cite{weber2010quantitative}, and is routinely recommended for tumor detection, stratification, therapy response assessment, and recurrence monitoring across multiple malignancies \cite{jadvar2017appropriate}.
Recent years have seen extensive use of 3D deep learning models, particularly U-Net-based architectures and their successors, for tumor segmentation on PET-CT scans \cite{wang2024dual,fallahpoor2024deep}. 3D networks capture volumetric context, enabling more consistent segmentation than 2D/2.5D methods \cite{gatidis2024results}. The U-Net family (and frameworks like nnU-Net) have become a de facto baseline for medical image segmentation due to their robust performance across tasks \cite{siddique2020u}.

A landmark effort addressing multi-cancer segmentation is the AutoPET challenge (MICCAI 2022), which introduced a large-scale whole-body FDG-PET/CT dataset comprising 1,014 studies for training \cite{gatidis2022whole,gatidis2023autopet}. The dataset included scans from 501 patients with histologically confirmed malignant melanoma, lymphoma, or non-small cell lung cancer, along with 513 negative controls, making it one of the most comprehensive multicancer PET/CT collections available \cite{gatidis2022fdg,jeblick2024psma}. To rigorously evaluate generalization, the AutoPET test set consisted of 150 scans acquired from two external institutions not represented in the training data \cite{gatidis2024results}.


3D nnU-Net has achieved Dice scores of 0.74–0.82 for lung tumors and ~0.70 for head-and-neck cases, outperforming inter-expert variability \cite{rokuss2024fdg,carles2024development}. However, performance declines (0.4–0.6) for fragmented diseases like lymphoma \cite{hasani2023automatic}. Robust generalization requires multi-institutional, multi-scanner, and multi-cancer training datasets, as shown in recent large-scale studies and challenges \cite{wang2025robust}. When such data are available, a 3D nnU-Net provides a strong baseline, often rivaling more complex models with minimal tuning \cite{isensee2024nnu}. Novel architectures, for example dual channel residual U-Nets, attention-enhanced fusion blocks, or semi-supervised multitask designs, have achieved modest improvements (Dice up to 0.87 in AutoPET III), but their gains remain incremental compared with the robustness and adaptability of nnU-Net \cite{wang2024dual,gatidis2023autopet}. To date, no study has systematically evaluated oesophageal cancer PET/CT for cross-domain tumor segmentation. We address this gap by introducing the first expert-annotated oesophageal cancer PET/CT dataset, alongside an independent lung cancer cohort, to stress-test cross-cancer generalizability of 3D nnU-Net.

AutoPET results revealed slight performance drops on the institution that was under-represented in training, indicating some domain shift effects \cite{gatidis2024results}. The challenge concluded that data quantity and quality had a larger impact on performance than architectural novelty.i.e. a well-trained 3D U-Net/nnU-Net with sufficient diverse data can generalize fairly well. The field is thus moving toward evaluating segmentation across multiple cancers, scanners, and populations.

In this paper, we stress-test generalization and clinical translation by introducing two novel datasets. Our contributions are:
\begin{itemize}
    \item We present the first whole-body oesophageal cancer PET/CT dataset (n=279) for segmentation research, together with an independent lung cancer dataset (n=54), enabling the first cross-cancer evaluation of nnU-Net.
    \item We benchmark 3D nnU-Net \cite{isensee2021nnu} generalization using both private datasets and the public AutoPET dataset across multi-center cohorts.
    \item We evaluate segmentation robustness across cancer types and populations, outlining key strengths and limitations for clinical translation.
\end{itemize}

\section{Methods}
\subsection{Dataset}

\begin{center}
    \begin{figure*}
        \centering
        \includegraphics[scale=0.400]{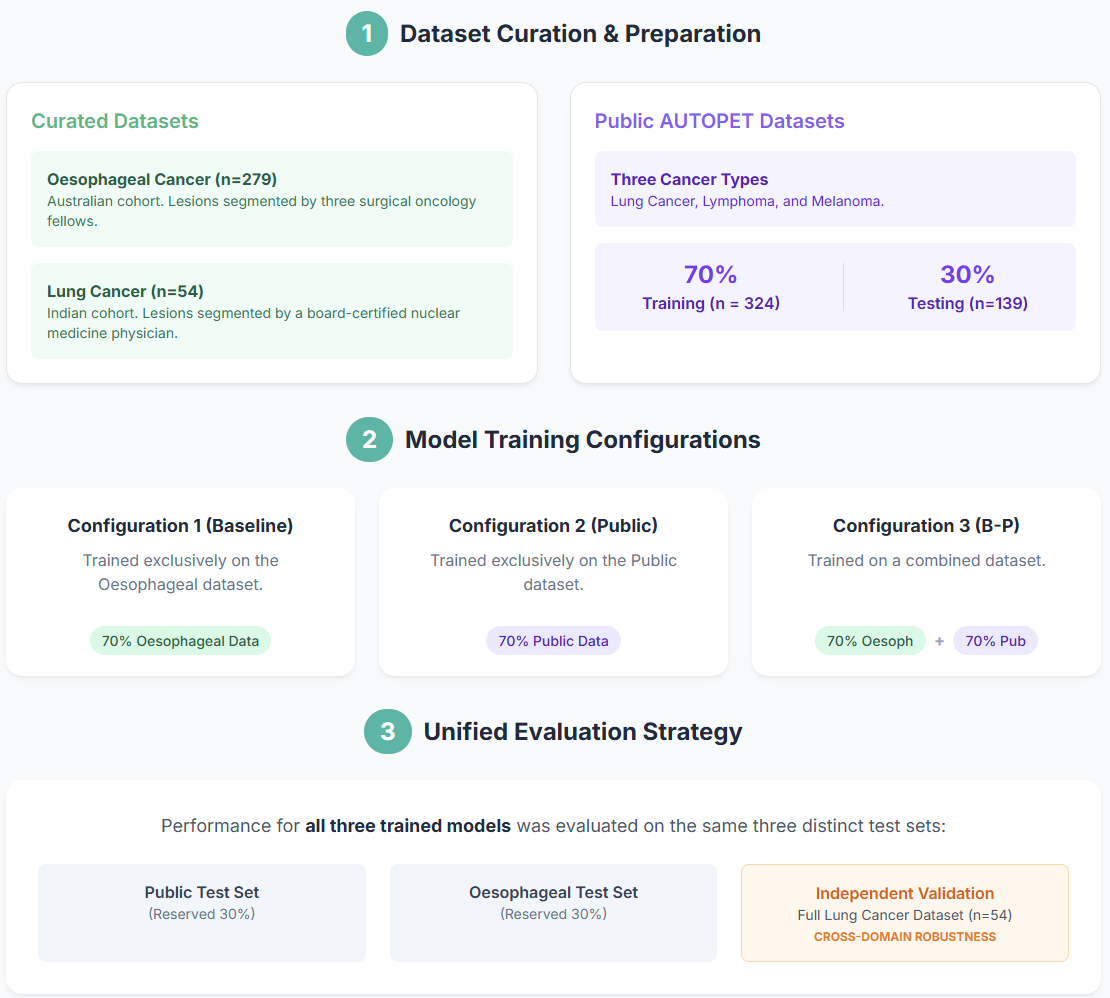}
        \caption{ Schematic visualizes the dataset curation, model training, and evaluation process of 3D nnU-Net. (Drawn in canvas using Gemini 2.5 pro)}
        \label{fig:F2}
    \end{figure*}
\end{center}

We curated an annotated data set for two types of cancers, oesophageal cancer (n = 279, Australian cohort) and lung cancer (n=54, Indian cohort both sybtypes; squamous cell carcinoma and adenocarcinoma). The Australian Oesophageal Cancer (OC) dataset also includes oesophageal adenocarcinoma and oesophageal squamous cell carcinoma. For the oesophageal cancer cohort, three surgical oncology fellows collaboratively performed manual segmentation of lesions, ensuring consensus-based annotation quality. Oesophageal cancer cohort study was approved the local human research ethics commitee.  For the lung cancer cohort, lesion segmentation was conducted by a board-certified nuclear medicine physician with clinical expertise in molecular imaging. This study has been approved by the local institutional ethics committee. All the annoataions were done using open source software ITK snap. In addition, we used publicly available AutoPET datasets comprising three distinct cancer types: lung cancer, lymphoma, and melanoma. The public datasets were randomly mixed  for the three types and partitioned into a 70:30 split, with 70 percent allocated for model training and 30 percent reserved for testing. Figure 2 describes the number of images used for training and testing in both curated and public configurations.

\begin{table}[]
\centering
\begin{tabular}{|ll|l|l|}
\hline
\multicolumn{2}{|c|}{\textbf{Datasets}}                                     & \multicolumn{1}{c|}{\textbf{Train}} & \multicolumn{1}{c|}{\textbf{Test}} \\ \hline
\multicolumn{1}{|l|}{\multirow{3}{*}{\textbf{Curated}}} & \textbf{Oesophagus (OC)}      & 210                                 & 69                                 \\ \cline{2-4} 
\multicolumn{1}{|l|}{}                                  & \textbf{Lung}  & 0                                   & 54                                 \\ \cline{2-4} 
\hline
\multicolumn{1}{|l|}{\textbf{Public}}                   & \textbf{Melanoma, Lung and Lymphoma (AutoPET)}  & 324                                 & 139                                \\ \hline
\end{tabular}
\caption{Overview of datasets used for model development and evaluation. The private datasets included OC (200 train, 69 test), Lung (54 test only). The public dataset AutoPET contributed 324 training cases and 139 test cases. Only OC and AutoPET were used for training, while all three datasets were used for independent evaluation to assess cross-dataset generalization.}
\end{table}

We trained a 3D nnU-Net segmentation model under three experimental configurations to evaluate generalisability and cross-cohort performance as shown in Figure \ref{fig:F2}. In the first configuration, the model was trained exclusively on the oesophageal cancer dataset (baseline), using a 70 percent training split. In the second configuration, the model was trained on 70 percent of the public dataset. In the third configuration, training was performed on a combined dataset comprising 70 percent of the public dataset and 70 percent of the oesophageal cancer dataset (combined training). Across all three training configurations, model performance was evaluated using the same validation strategy: testing on the reserved 30 percent portion of the public dataset and on the oesophageal cancer test set, with additional independent validation on the lung cancer dataset to assess cross-domain robustness. 

The demographic analysis (Figure 2, top panel) highlights notable differences between the Australian oesophageal cancer cohort (n = 279) and the Indian lung cancer cohort (n = 54). The lung cancer cohort, by contrast, demonstrates a comparatively narrower range, particularly for height and age at diagnosis. Notably, the mean height appears higher in the Australian cohort, consistent with broader anthropometric differences between populations, while median age at diagnosis trends older in the lung cancer group, potentially reflecting differences in disease epidemiology and healthcare-seeking patterns.

\begin{center}
    \begin{figure*}
        \centering
        \includegraphics[scale=0.18]{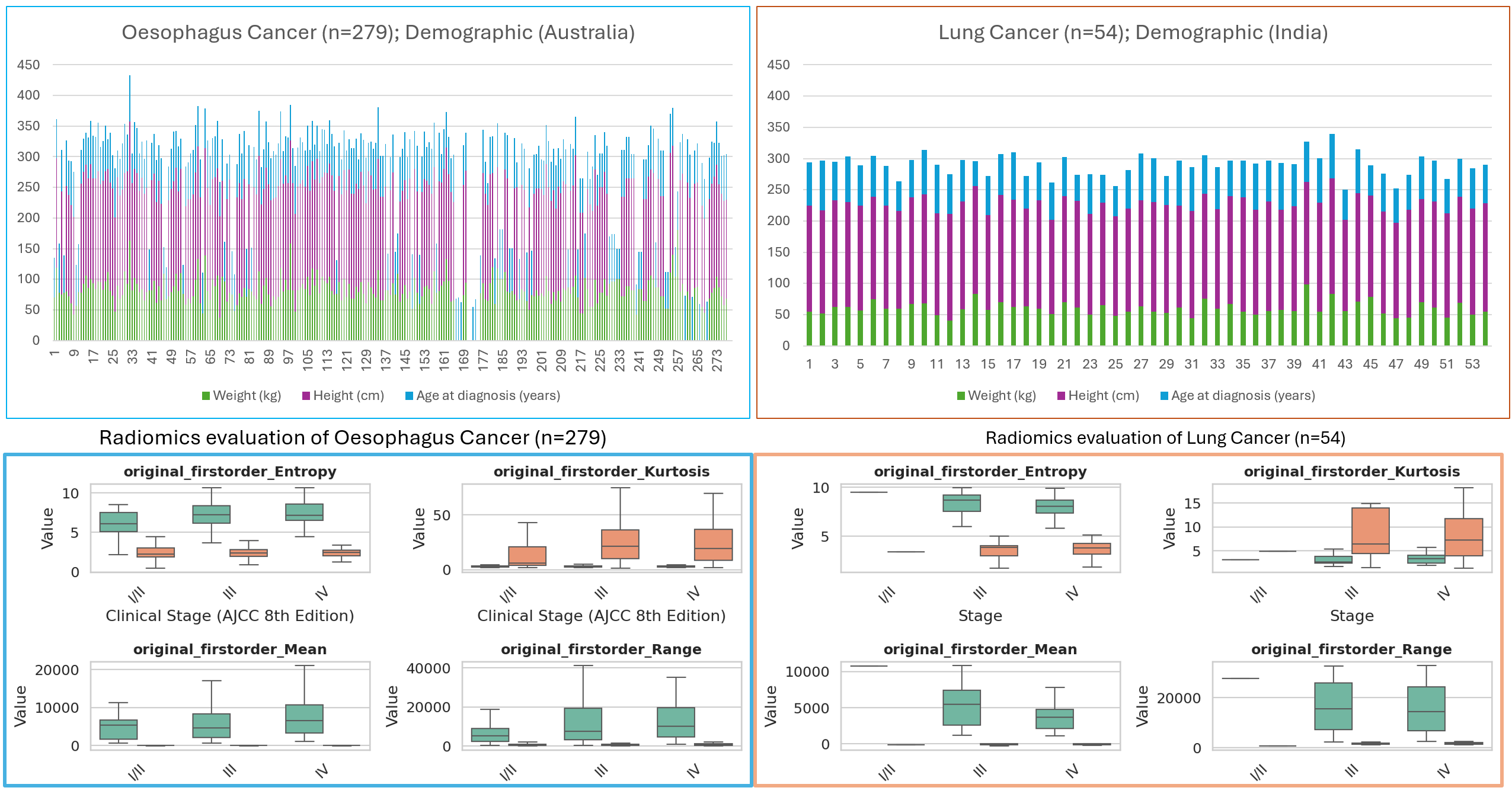}
        \caption{Representation of the datasets (green: PET, orange: CT).}
        \label{fig:F1}
    \end{figure*}
\end{center}

Radiomics features are computed using Pyradiomics package \cite{van2017computational}. Its evaluation (Figure \ref{fig:F1}, bottom panel) further distinguishes the two cohorts. For oesophageal cancer, PET-derived features such as Entropy and Mean display pronounced stage-dependent variation, with higher entropy and mean values in advanced stages (III–IV), indicating increased lesion heterogeneity and uptake intensity. CT-derived features (e.g., Kurtosis and Range) also increase with clinical stage, though the variance across stages is less marked than for PET metrics, suggesting complementary but distinct diagnostic sensitivity. In the lung cancer cohort, PET-derived entropy and range again increase with clinical stage, but kurtosis shows a more dramatic stage-related rise compared to the oesophageal cohort, possibly reflecting sharper intensity distributions within lesions. Additionally, PET-derived mean values for lung cancer tend to be lower overall than for oesophageal cancer, while range values are higher, hinting at differing lesion uptake patterns and spatial intensity variability between the malignancies.

\subsection{Preprocessing, nnU-Net Architecture and Configuration}
All scans were resampled to 1\texttimes1\texttimes1 mm\textsuperscript{3}. CT intensities were clipped to [--1000, 1000] HU and normalized. PET intensities were directly scaled.

For this study, we utilized nnU-Net v2, chosen for its self-configuring nature that dynamically determines optimal patch sizes, network architectures, and training schedules. Specifically, we employed a 3D full-resolution U-Net architecture featuring deep supervision, instance normalization, and LeakyReLU activation functions.

The model was trained for 250 epochs using a patch size of 128x128x128, a batch size of 2, and a combined Dice and Cross-Entropy loss function. Optimization was performed using Stochastic Gradient Descent (SGD) with Nesterov momentum (mu=0.99), starting with a learning rate of 0.01 that followed a polynomial decay schedule. To prevent overfitting, an early stopping mechanism was implemented with a patience of 50 epochs. Default augmentations (rotation, scaling, elastic deformation, gamma correction) were applied. Postprocessing removed small spurious objects. we have used two RTX A6000 NVIDIA GPU card with 24 GB memory each for training.

\subsection{Evaluation Metrics}
Performance metrics included the Dice Similarity Coefficient (DSC), precision, recall and Hausdorff Distance at the 95th percentile (HD95) \cite{muller2022towards}, providing a comprehensive assessment of segmentation accuracy, lesion boundary delineation and spatial agreement with expert annotations.

\section{Results and Analysis}

\subsection{Quantitative Analysis}
Table \ref{tab:results} summarizes nnU-Net performance when trained on OC, AutoPET, or combined datasets and tested across Lung, OC, and AutoPET cohorts.
Models trained on single datasets showed strong in-domain performance but limited generalization. The OC-only model achieved the best results on its own test set (DSC = 57.8, Precision = 62.2, HD95 = 3.7) but collapsed when evaluated on lung or AutoPET cohorts (DSC $\leq$ 3.4), highlighting severe cancer-type and site-specific overfitting. In contrast, the AutoPET model generalized better, performing strongly on its native domain (DSC = 63.5, Recall = 71.5) and moderately on external cohorts, though with reduced precision on OC (DSC= 26.7). The combined OC+AutoPET model delivered the most balanced results across all domains (mean DSC = lung (52.9); oesophageal (40.7); AutoPET (60.9)), with lower HD95 values, confirming the value of multi-center training for robust segmentation.

\begin{table}[]
\centering
\scalebox{0.7}{
\begin{tabular}{|l|ccc|ccc|ccc|ccc|}
\hline
\multirow{2}{*}{\textbf{Models}} & \multicolumn{3}{c|}{\textbf{DSC}} & \multicolumn{3}{c|}{\textbf{Precision}} & \multicolumn{3}{c|}{\textbf{Recall}} & \multicolumn{3}{c|}{\textbf{HD95}} \\ \cline{2-13} 
 & \textbf{Lung} & \textbf{OC} & \textbf{AutoPET} & \textbf{Lung} & \textbf{OC} & \textbf{AutoPET} & \textbf{Lung} & \textbf{OC} & \textbf{AutoPET} & \textbf{Lung} & \textbf{OC} & \textbf{AutoPET} \\ \hline
\textbf{OC}          & 1.25 & 57.8 & 3.4  & 11.1 & 62.2 & 34.1  & 0.8  & 64.5 & 2.1  & 42.0  & 3.7  & 99.5  \\ \hline
\textbf{AutoPET}      & 51.6 & 26.7 & 63.5 & 43.6 & 17.5 & 67.8  & 80.8 & 77.9 & 71.5 & 105.7 & 81.3 & 60.7  \\ \hline
\textbf{OC+AutoPET}  & 52.9 & 40.7 & 60.9 & 45.7 & 35.6 & 66.0  & 75.8 & 62.9 & 68.8 & 72.9  & 59.5 & 67.7  \\ \hline
\end{tabular}
}
\caption{Cross-dataset performance of nnU-Net models trained on OC, AutoPET, and combined (OC+AutoPET). Metrics include DSC, Precision, Recall, and HD95.}
\label{tab:results}
\end{table}

The results demonstrate a clear trade-off between domain specificity and generalization. Models trained on single-domain data achieve high in-domain accuracy but fail to transfer across cohorts, whereas training solely on public heterogeneous data enables broader generalization but underperforms on oesophageal cancer. In contrast, combined training provides the most balanced outcome, delivering stable segmentation across heterogeneous cohorts with improved boundary accuracy. These findings provide quantitative evidence that dataset diversity, particularly multicenter and multicancer integration, is a critical determinant of clinically robust and generalizable segmentation models.


\begin{center}
    \begin{figure*}
        \centering
        \includegraphics[scale=0.3]{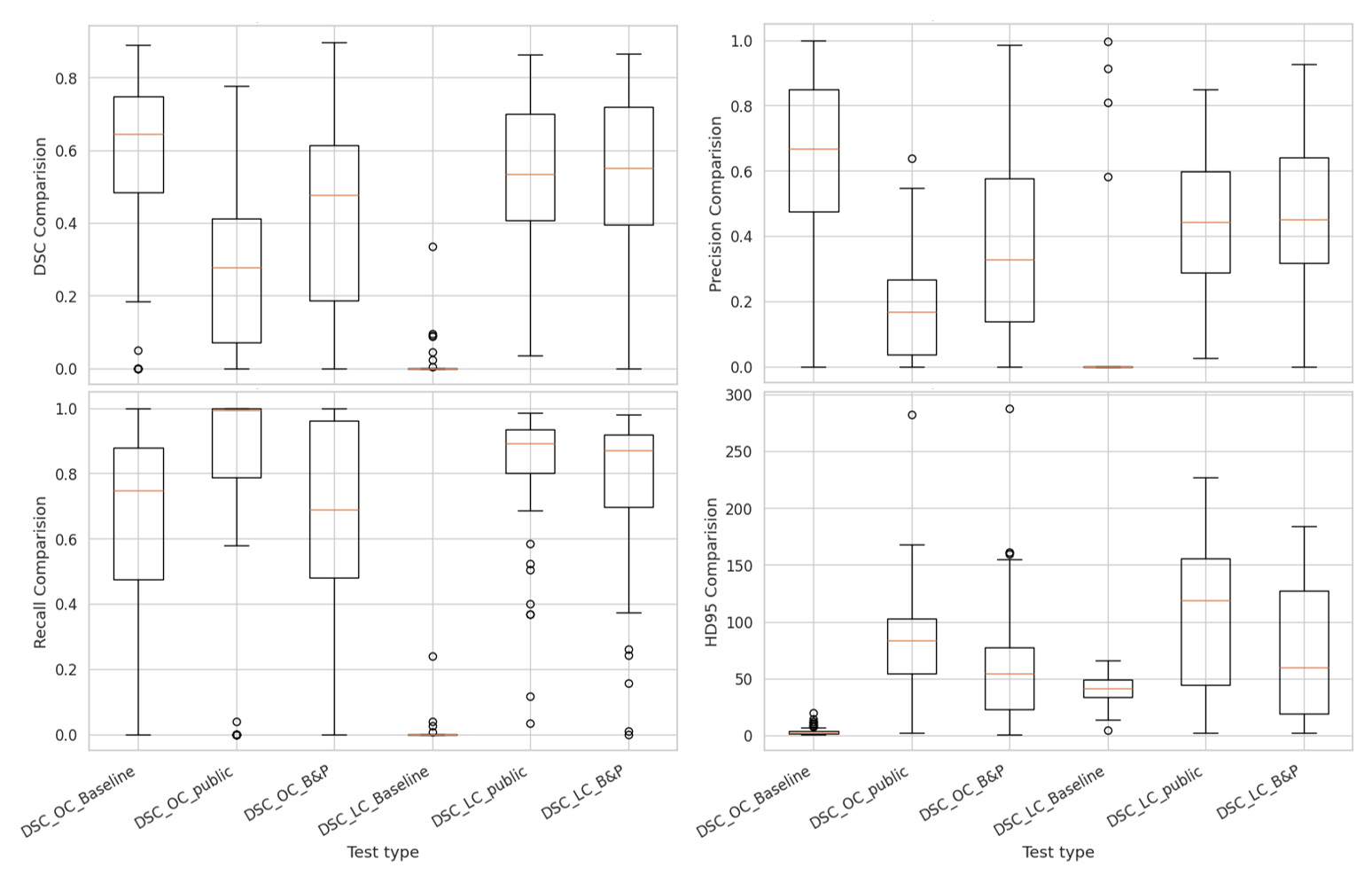}
        \caption{Boxplots of segmentation performance (DSC, Precision, Recall, HD95) across oesophageal cancer (OC) and lung cancer (LC) test cohorts. Models were trained on OC-only (baseline), public-only, and combined (baseline+public) datasets. The baseline achieves strong in-domain performance but fails cross-domain, the public model generalizes better but underperforms on OC, while combined training provides the most balanced results.}

        \label{fig:F5}
    \end{figure*}
\end{center}
Figure~\ref{fig:F5} presents boxplots of DSC, Precision, Recall, and HD95 across oesophageal cancer (OC) and lung cancer (LC) test sets under three training configurations: baseline (OC), public-only (AutoPET), and combined (baseline+public). The OC baseline achieves the best within-domain accuracy with high DSC, Precision, and low HD95, but fails completely on lung cancer, reflecting severe site-specific overfitting. Conversely, the public-only model generalizes better across cohorts, particularly in lung cancer, but underperforms on oesophageal cancer with reduced boundary accuracy. The combined model offers the most stable compromise, maintaining higher DSC and Recall across both cohorts, albeit with slightly higher HD95 than the in-domain baseline.

\subsection{Qualitative Analysis}
\begin{center}
    \begin{figure*}
        \centering
        \includegraphics[scale=0.18]{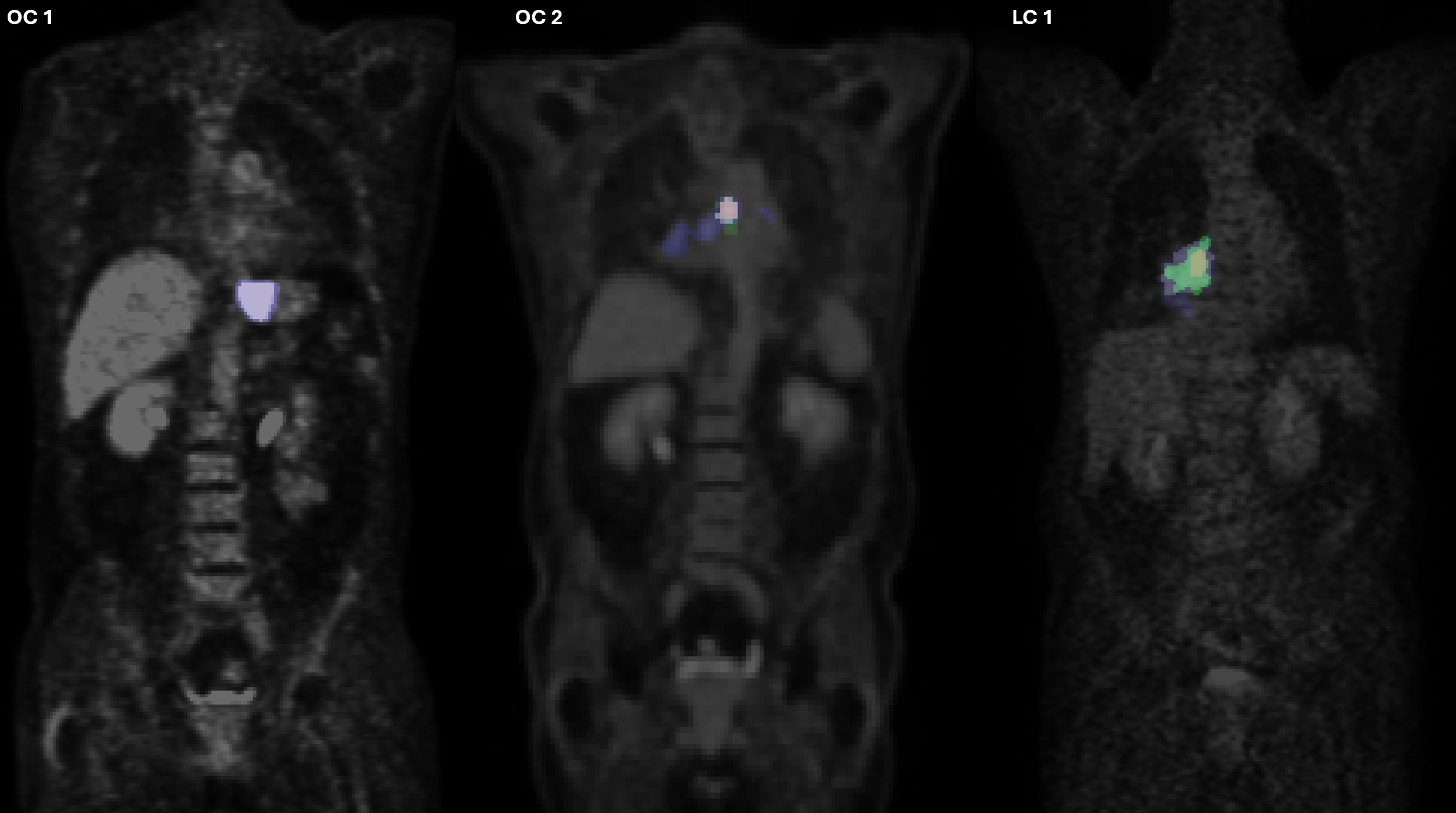}
        \caption{Coronal PET-CT views present a whole-body comparison of segmentation outcomes for two oesophageal cancer test cases (OC 1 and OC 2) and one independent lung cancer test case (LC 1) across three training configurations: Baseline (target-domain only), Public-only, and combined (public + target-domain). The ground truth contours are shown in green, while the overlays of the baseline, public-only, and combined models appear in red, blue, and gray, respectively.}
        \label{fig:F3}
    \end{figure*}
\end{center}
\begin{center}
    \begin{figure*}
        \centering
        \includegraphics[scale=0.18]{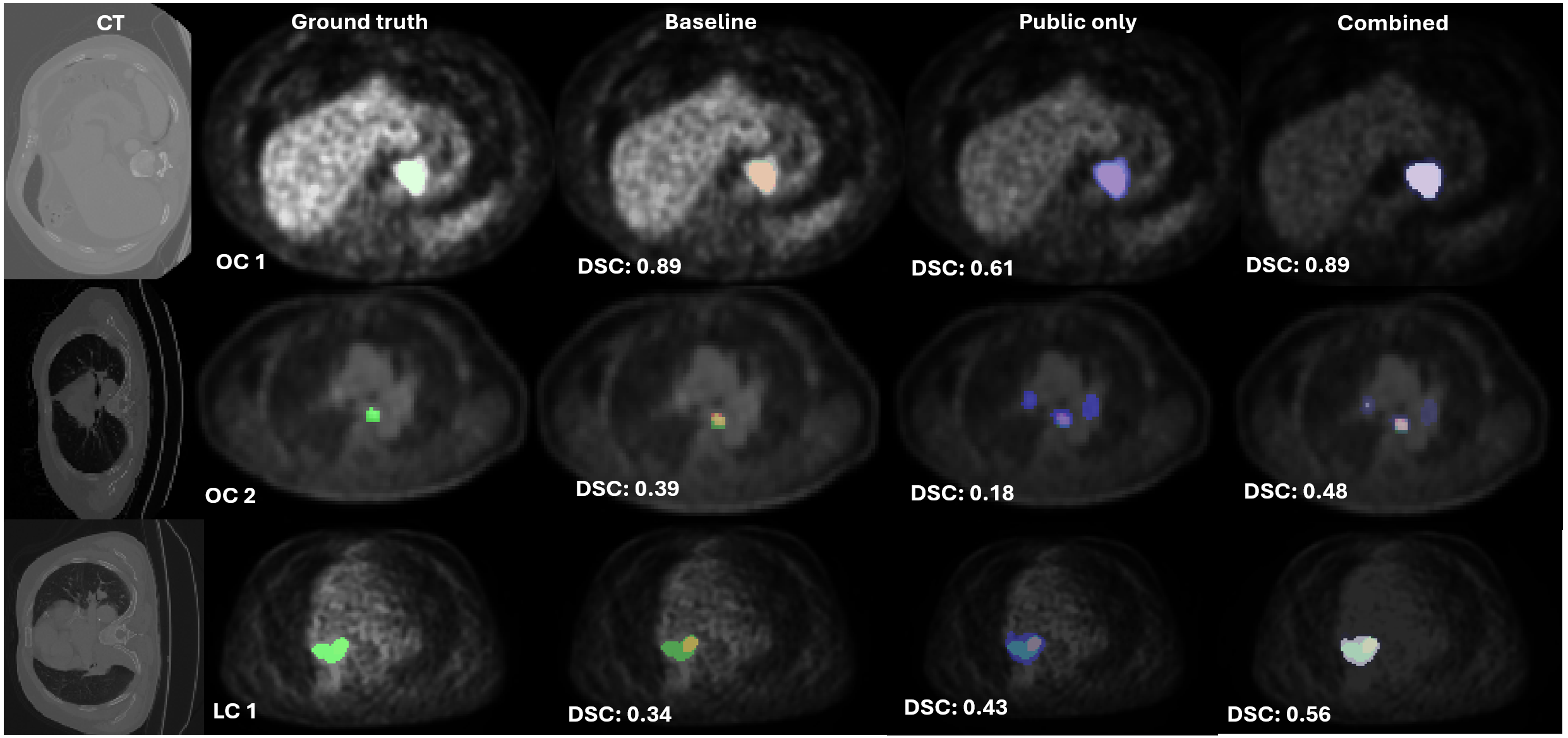}
        \caption{The axial PET-CT slices in the figure illustrate qualitative and quantitative differences in segmentation performance across three training configurations}
        \label{fig:F4}
    \end{figure*}
\end{center}

The coronal PET/CT views in Figure \ref{fig:F3} present a whole-body comparison of segmentation outcomes for two oesophageal cancer test cases (OC 1 and OC 2) and one independent lung cancer test case (LC 1) across three training configurations: Baseline (target-domain only), Public-only, and combined (baseline - public) domain. In axial view as shown in Figure \ref{fig:F4} the ground truth contours are shown in green, while overlays from the Baseline, Public-only, and combined models appear in red, blue, and gray, respectively.

For OC 1, which represents a large, high-contrast oesophageal lesion, both the Baseline and combined models achieve near-perfect volumetric delineation (DSC = 0.89), capturing the lesion’s boundaries without over- or under-segmentation. The Public-only model, while correctly identifying the lesion core, underestimates its full extent, resulting in a reduced DSC of 0.61 and overestimating peripheral uptake regions. This underperformance reflects the loss of disease-specific sensitivity when training excludes oesophageal-specific data.

OC 2, characterised by multiple small, low-contrast lesions, proves challenging for all models. The Baseline configuration yields sparse and fragmented detections (DSC = 0.39), while the Public-only model performs worst (DSC = 0.18), producing small, scattered predictions with limited coverage. The combined configuration improves capture modestly (DSC = 0.39), recovering more lesion volume but still failing to produce complete, contiguous segmentations. These findings illustrate the limitations of all three models in handling low intensity, small-volume lesions where contrast and morphology are less distinct.

In LC 1, which contains a moderately sized lung lesion, the Baseline (oesophageal trained) model achieves only partial coverage (DSC = 0.34), indicating limited cross-cancer transferability. Public-only training improves the delineation (DSC = 0.43) by leveraging diverse cancer representations in the training set. The combined approach delivers the best performance (DSC = 0.56), demonstrating the benefit of combining domain-specific and heterogeneous datasets.

\subsection{Generalizability across heterogeneous location}

Generalizability across anatomically and metabolically heterogeneous locations in PET-CT segmentation tasks remains a core challenge, as evidenced by the comparative results across oesophageal and lung cancer test sets. Models trained solely on a single cancer type (baseline) exhibit high specificity and accurate lesion delineation within their native domain, particularly for lesions with high contrast and well-defined boundaries. However, when applied to anatomically distinct sites, such as transferring oesophageal-trained models to lung lesions, performance drops significantly due to differences in lesion morphology and background uptake distribution. This pattern reflects a narrow learned feature space, finely tuned to the training cohort’s uptake characteristics but lacking the variability necessary to adapt to divergent anatomical contexts.

In contrast, training exclusively on a heterogeneous public dataset expands the model’s exposure to varied lesion locations and uptake profiles, improving cross-site adaptability most notably seen in improved lung cancer segmentation compared to baseline. However, this gain in generalisability comes at the expense of precision within the original domain, where specialised metabolic and anatomical cues are diluted by broader, less disease-specific representations. The combined training strategy partially resolves this trade-off, leveraging the disease-specific sensitivity of target-domain data while retaining the adaptability learned from diverse public cases. This integration yields more stable performance across anatomically disparate sites, although performance remains constrained for small, low-contrast lesions, indicating that variability in lesion location alone is insufficient without targeted enhancement of sensitivity to subtle uptake patterns.


\subsection{Error Analysis of Poorly Performing Subjects}
Several test subjects exhibited very low DSC ($\leq$ 0.4), with multiple cases scoring zero. Careful review identified several recurring issues that likely contributed to these failures. First, very small lesions proved challenging for nnU-Net, consistent with known difficulties in segmenting small targets due to partial volume effects in PET and limited contrast. Second, mismatched or PET-CT scans were suspected in some cases, where PET uptake did not align with the CT anatomy, possibly due to wrong scan pairing (clinical dataset curation limitations). Third, in some cases, the lesion was not clearly visible on PET or the algorithm mistakenly segmented surrounding structures such as Barrett’s esophagus or the entire stomach instead of the tumour. Such edge cases underscore the importance of careful data curation and highlight scenarios where automated segmentation may require manual review or exclusion.

\section{Discussion}
This study provides the first cross-cancer evaluation of nnU-Net on oesophageal PET/CT, addressing a gap in prior segmentation work that has largely focused on lung, head-and-neck, melanoma, prostate and lymphoma. Unlike previous studies that optimize nnU-Net for a single cancer type, we deliberately evaluated performance under domain shift using oesophageal cancer, lung cancer, and the heterogeneous AutoPET dataset. 

The results indicates that dataset diversity outweighs architectural novelty: single-domain training yields strong within-cohort performance but fails to transfer, while public-only training generalizes better but loses precision on oesophageal cancer. The combined strategy strikes the most practical balance, supporting robust performance across heterogeneous cohorts. These findings align with emerging evidence that even bespoke architectures (attention- or fusion-based) provide only marginal gains (1–5 DSC points) over a well-trained nnU-Net, underscoring the primacy of multi-center, multi-cancer data for clinical deployment.

From a translational perspective, balanced cross-cancer performance is essential in real-world workflows where diverse anatomical and metabolic contexts are routinely encountered. Our error analysis further highlights limitations in segmenting small, low-contrast lesions, misaligned scans, and atypical postoperative anatomies, where automated methods remain unreliable. Such edge cases reinforce the importance of human-in-the-loop integration \cite{vashistha2025potential}, for example through interactive annotation platforms like MONAI Label, which enable rapid correction and continual adaptation.

\section{Conclusion}
This study demonstrates that combining diverse public datasets with domain-specific PET-CT data provides the most balanced trade-off between in-domain accuracy and cross-domain generalizability for multi-cancer tumor segmentation using nnU-Net. The findings highlight that dataset diversity, particularly multi-center and multi-cancer integration, rather than architectural novelty, is the primary driver of robust and clinically relevant performance. Nonetheless, persistent challenges with small, low-contrast lesions emphasize the importance of embedding such models within human-in-the-loop systems to ensure reliability and safe clinical translation.

\section*{Acknowledgments}

The authors wish to thank the patients that have participated in our research project. We acknowledge the work and support of the Upper GI Unit at the Princess Alexandra Hospital, Brisbane, Australia. This research was partially carried out at the Translational Research Institute, Woolloongabba, QLD 4102, Australia. The Translational Research Institute is supported by a grant from the Australian Government. This study was funded by a large Cancer Council Queensland (CCQ) grant, aimed at Accelerating Collaborative Cancer Research. Lauren Aoude is supported by a National Health and Medical Research Council (NHMRC) Emerging Leadership 2 (EL2) Grant (APP2034399). Rajat Vashistha acknowledges his fellowship from CCQ ACCR-190. Christine Jestin Hannan would also acknowledge a grant from Bengt Ihres Foundation.

\section*{Declaration of generative AI and AI-assisted technologies in the writing process}
During the preparation of this work the author(s) used Gemini 2.5 pro in order to improve the grammatical correction and readability. After using this tool/service, the author(s) reviewed and edited the content as needed and take(s) full responsibility for the content of the publication

\bibliographystyle{plain}
\bibliography{biblo}

\end{document}